\documentclass[fleqn,usenatbib]{mnras}

\usepackage{newtxtext,newtxmath}

\usepackage[T1]{fontenc}
\usepackage{ae,aecompl}

\usepackage{graphicx}	
\usepackage{amsmath}	
\usepackage{amssymb}	
\usepackage{subfigure}
\usepackage{mathtools}
\usepackage{footmisc}
\DeclarePairedDelimiter{\ceil}{\lceil}{\rceil}
\title[SpecPhot]{SpecPhot: A Comparison of Spectroscopic and Photometric Exoplanet Follow-Up Methods}

\author[B. F. Cooke \& D. Pollacco]{
Benjamin F. Cooke$^{1,2}$\thanks{E-mail: b.cooke@warwick.ac.uk}
and Don Pollacco$^{1,2}$
\\
$^{1}$Department of Physics, University of Warwick, Gibbet Hill Road, Coventry CV4 7AL, UK\\
$^{2}$Centre for Exoplanets and Habitability, University of Warwick, Gibbet Hill Road, Coventry CV4 7AL, UK\\
}

\date{Accepted XXX. Received YYY; in original form ZZZ}

\pubyear{2019}

\begin{document}
\label{firstpage}
\pagerange{\pageref{firstpage}--\pageref{lastpage}}
\maketitle

\begin{abstract}
We set out a simulation to explore the follow-up of exoplanet candidates. We look at comparing photometric (transit method) and spectroscopic (Doppler shift method) techniques using 
three instruments: NGTS, 
HARPS and CORALIE. We take into account precision of follow-up and required observing time in attempt to rank each method for a given set of planetary system parameters. The methods are assessed on two criteria, SNR of the detection and follow-up time before characterisation. We find that different follow-up techniques are preferred for different regions of parameter space. For SNR we find that the ratio of spectroscopic to photometric SNR for a given system goes like $R_p/P^{\frac{1}{3}}$. For follow-up time we find that photometry is favoured for the shortest period systems ($<10$\,d) as well as systems with small planet radii. Spectroscopy is then preferred for systems with larger radius, and thus more massive, planets (given our assumed mass-radius relationship). Finally, we attempt to account for availability of telescopes and weight the two methods accordingly.
\end{abstract}




\section{Introduction}
\label{sec:Introduction}

Exoplanets are being discovered in larger numbers than ever before. 
Many of these new planets are being detected from the large data sets produced by long photometric campaigns. Thus it is more important than ever to have an efficient plan to best utilise follow-up efforts. As an added impact of these surveys we are beginning to reach into longer period parameter space. Unfortunately, due to the functionality of these photometric surveys, longer period systems have an increased chance of only exhibiting a single transit within a survey leading to significant yields of monotransits \citep[e.g. ][]{2018A&A...619A.175C,2019AJ....157...84V}. Therefore these systems have only estimated periods based the shape of the discovery transit \citep{2016MNRAS.457.2273O}. To properly characterise these, and other, systems we need to follow them up. Generally we can follow-up exoplanets using either photometric or spectroscopic techniques.


Photometry as an exoplanet detection method relies on the transit method. That is, the reduction in flux from a host star as an exoplanet passes across an observers line of sight. Assuming edge on systems the size of the dip then gives the radius ratio of the star-planet system. Notable photometry missions include the Wide Angle Search for Planets \citep[WASP,][]{2006PASP..118.1407P}, Kepler/K2 \citep{2010Sci...327..977B,2014PASP..126..398H}, the Next-Generation Transit Survey \citep[NGTS,][]{2018MNRAS.475.4476W} and the Transiting Exoplanet Survey Satellite \citep[TESS,][]{2015JATIS...1a4003R}.

The first detection of a transiting exoplanet was HD 209458b in 1999 \citep{2000ApJ...529L..45C,2000ApJ...529L..41H} (previously discovered using radial velocities) and the first exoplanet discovered via this method was OGLE-TR-56b in 2002 \citep{2002AcA....52..115U}. Photometry can be carried out from space or ground-based with both methods having pros and cons. This paper focuses on ground-based photometric follow-up. Generally space based photometry can detect smaller planetary signals due to reduced instrumental noise, a lack of atmospheric corrections and more consistent sampling but ground-based photometry is improving constantly. The smallest signal discovered from the ground to date is 0.13\% \citep{2019MNRAS.486.5094W}. As of November 2019 3150/4093 ($\sim$\,78\%) of confirmed exoplanets have been discovered via this method\footnote{\href{https://exoplanetarchive.ipac.caltech.edu/}{https://exoplanetarchive.ipac.caltech.edu}\label{fn:repeat}}.

Spectroscopy uses the radial velocity or Doppler shift technique to detect exoplanets. As the star and planet system orbit a common centre of mass the light from the star is red/blue-shifted along the observers line of sight. Measuring this periodic shift can reveal the presence of an exoplanet. Assuming edge-on, circularly orbiting systems, the size of the motion then gives the ratio of planet mass to total system mass to the power 2/3. Notable spectroscopic facilities include the High-Accuracy Radial-velocity Planetary Search \citep[HARPS,][]{2003Msngr.114...20M}, CORALIE \citep{2000A&A...354...99Q}, and the Echelle SPectrograph for Rocky Exoplanets and Stable Spectroscopic Observations \citep[ESPRESSO,][]{2014AN....335....8P}.

The first planet discovered via this method was 51 Pegasi b in 1995, which was also the first exoplanet to be found orbiting a sun-like star \citep{1995Natur.378..355M}. As of November 2019 779/4093 ($\sim$\,19\%) of confirmed exoplanets have been discovered via this method\footnote{\href{https://exoplanetarchive.ipac.caltech.edu/}{https://exoplanetarchive.ipac.caltech.edu}}.



Both photometry and spectroscopy require significant time and resources to properly follow-up exoplanet systems. As such it is vital to not waste telescope resources by using an inefficient technique when a different one could be carried out much more cheaply in terms of competitive time. To ensure this it is necessary to know which areas of parameter space lend themselves most naturally to each method. There are two criteria that are notable here, namely instrumental SNR (the reliability with which an instrument can detect a signal of a given magnitude) and follow-up time (how much time would be required for a given technique to successfully follow-up an exoplanet).

In this paper we attempt to answer these questions. We set out our method to choose the best follow-up procedure for an exoplanet that has discovery photometry (i.e. is transiting) but only an estimated or unsure period (a monotransit for example). This is the situation for many systems discovered as part of large photometric surveys. Section \ref{sec:Methodology} discusses our methodology, sets out the instruments and methods used and our SNR and follow-up time definitions and criteria. Section \ref{sec:Results} lays out our results and sections \ref{sec:Discussion} and \ref{sec:Conclusions} give our discussions and the projects conclusions.

\section{Methodology}
\label{sec:Methodology}

To attempt to determine the feasibility of using each method for follow-up of exoplanet candidates we first defined a grid of points covering a range of exoplanetary parameter space. It was decided to define a point by 3 parameters as this lends itself easily to 3D representations of results. Additionally, using a small number of parameters increase the generality of the results. The first parameter chosen is stellar radius. Since we are interested in the follow-up of planetary systems with a discovery transit we assume that stellar radius will be a known parameter for the majority of systems as a result of comprehensive input catalogues for photometric surveys (e.g. the Kepler Input Catalogue \citep[KIC,][]{2011AJ....142..112B} and the TESS Input Catalogue \citep[TIC,][]{2019AJ....158..138S}). Our second chosen parameter is planetary radius as this can be inferred from the discovery transit and the stellar radius. Finally we choose period as our third parameter due to the strong dependence of follow-up efforts on period. Additionally, period is a parameter that can be estimated from the discovery transit \citep{2016MNRAS.457.2273O}.

The range of parameters chosen were determined from looking at a set of exoplanet parameters drawn from the NASA exoplanet archive \citep{2013PASP..125..989A}. The chosen ranges were $0.1 \leq R_\star \leq 10.0\,R_\odot$, $0.1 \leq R_p \leq 2.0\,R_{Jup}$ and $1.0 \leq P \leq 1000$\,days. The simulation could then be carried out on every point in the grid of parameter space and a comparison of the two follow-up methods could be made. We create the parameter grid using 40 points along each parameter axis evenly separated in log-space. For each point in this parameter space two values needed to be determined. Firstly, the possibility that a signal corresponding to a point with these parameters could be detected using either of the two methods. This was defined as the possibility that the signal could be detected with $SNR \geq 3.0$ for a given instrument. Secondly, it needed to be calculated how long it would take before a point with these parameters could be effectively followed up using either of the two methods.

\subsection{SNR}
\label{sec:SNR}

To calculate the SNR we first needed to calculate the size of the relevant signal based on the three known parameters, $R_\star$, $R_p$ and $P$. For photometry this is trivial if we assume the system is edge on to the observer (a reasonable assumption as we are only concerned with systems that have already be found to transit). The measurable signal is then simply the fractional reduction in flux caused by a transit;

\begin{equation}
\label{eq:phot signal}
    \delta = \left(\frac{R_p}{R_\star}\right)^2,
\end{equation}

where $\delta$ is the size of the photometry signal \citep{2010arXiv1001.2010W}.

For spectroscopy the process is a little more involved. Under the assumption of an edge on, circular orbit the equation for radial velocity amplitude (the signal that must be detected) is given by;

\begin{equation}
\label{eq:spec signal}
    K = \frac{2 \pi a M_p}{\left(M_\star + M_p\right) P},
\end{equation}

where $K$ is the RV semi amplitude, $a$ is the semi-major axis and $M_\star$ and $M_p$ are the masses of the star and planet respectively \citep{2010exop.book...27L}. The semi-major axis is given by

\begin{equation}
\label{eq:a}
    a = \left(\frac{P^2 G \left(M_\star + M_p\right)}{4 \pi^2}\right)^{\frac{1}{3}},
\end{equation}

where $G=6.67408\times10^{-11}\mathrm{m^3kg^{-1}s^{-2}}$ is the gravitational constant \citep{2010exop.book...27L}.

Calculating the two masses requires finding a mass-radius relationship for both stars and planets.

\subsubsection{Mass-radius relations}
\label{sec:Mass-radius relation}

\begin{figure}

    \begin{subfigure}[Stellar data
    \label{fig:mass-radius stars}]
    {\includegraphics[width=\columnwidth]{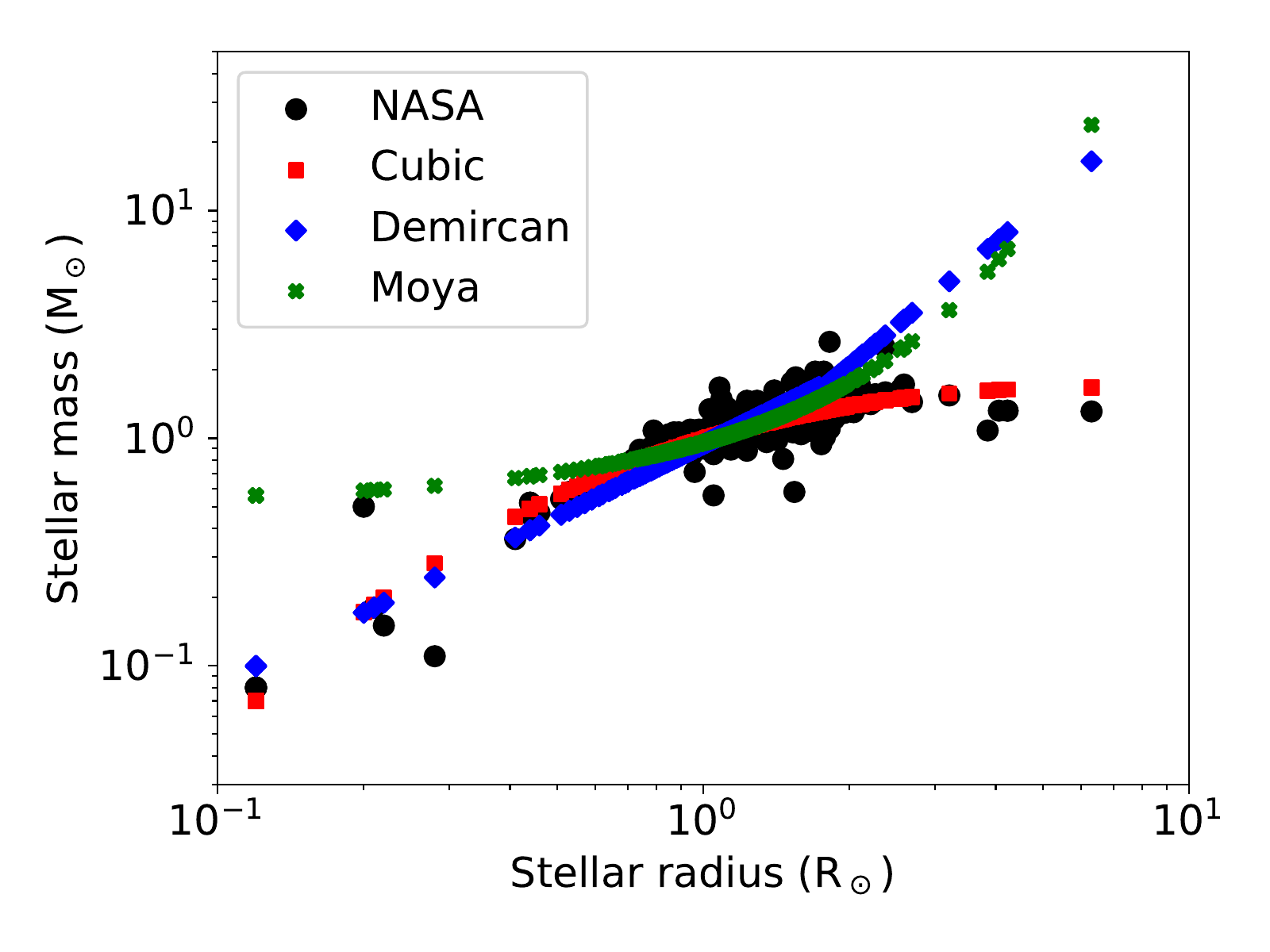}}
    \end{subfigure}
    \begin{subfigure}[Planetary data
    \label{fig:mass-radius planets}]
    {\includegraphics[width=\columnwidth]{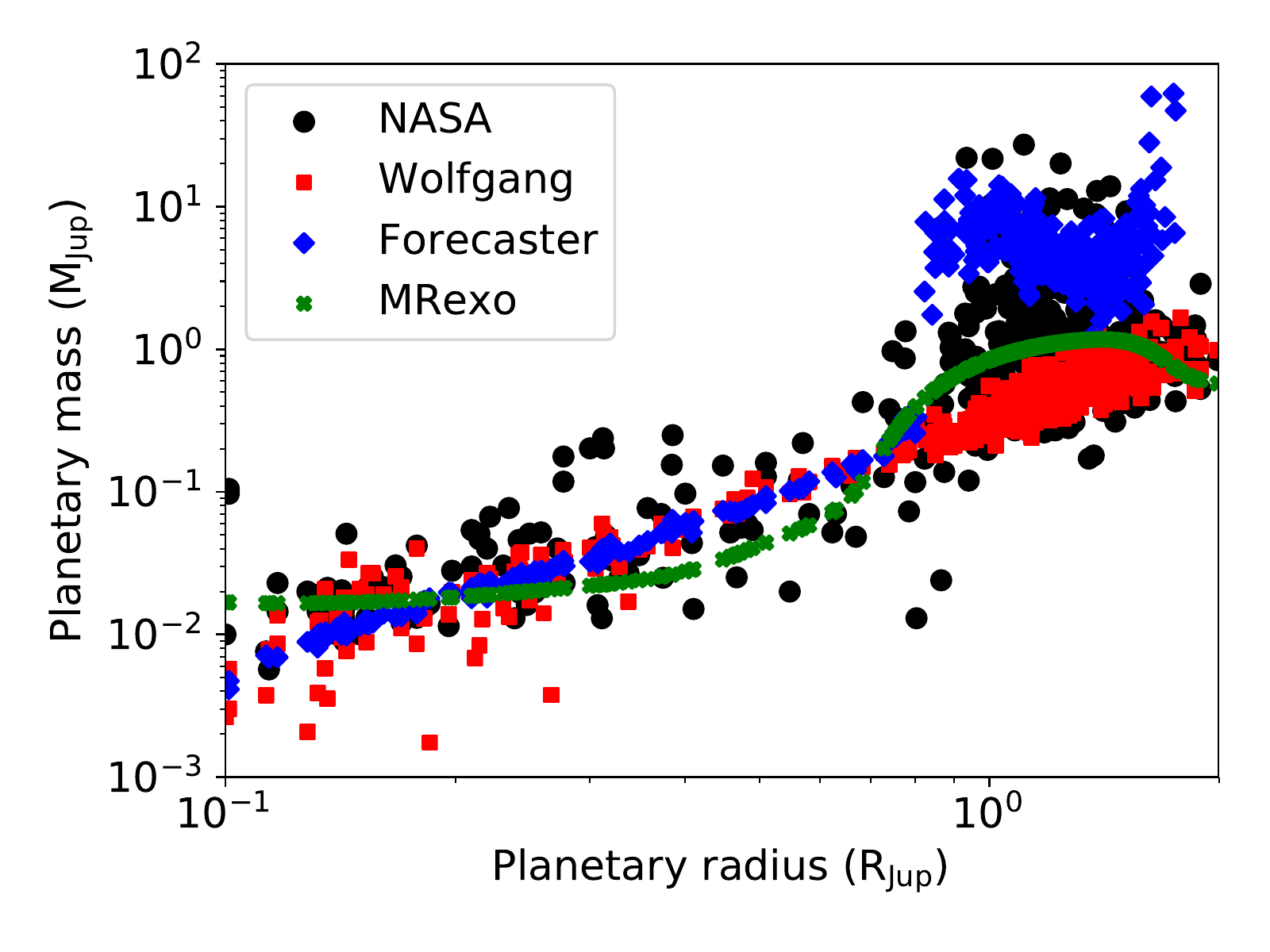}}
    \end{subfigure}
    \caption{Mass radius relations, NASA data is in black.\newline
    (a) Stellar relations, the colours are as follows; Red squares: cubic fit, Blue diamonds: Demircan \citep{1991Ap&SS.181..313D}, Green crosses: Moya \citep{2018ApJS..237...21M}.\newline
    (b) Planetary relations, the colours are as follows; Red squares: Wolfgang \citep{2016ApJ...825...19W}, Blue diamonds: Forecaster \citep{2017ApJ...834...17C}, Green crosses: MRExo \citep{2019arXiv190300042K}.}

\label{fig:mass-radius}
\end{figure}

To determine the best mass-radius relation for stars and planets a set of data was taken from the NASA exoplanet archive. Since this paper focuses on the follow-up of planetary systems already found to transit the planetary data used include only transiting planets. The stellar data however, used all planet hosts from the exoplanet archive. These data were then filtered leaving only those data points for which mass and radius are known to a fractional uncertainty of 30\% or better. To determine a mass-radius relation based on these data multiple techniques were tested.
Figures \ref{fig:mass-radius stars} and \ref{fig:mass-radius planets} shows the results of these efforts. The NASA data is in black with the different colours corresponding to the range of different fits and relations that were tested.

For the stellar relationship the plots show the results of a cubic fit (in log-log space) and two published relations; \cite{1991Ap&SS.181..313D} (a one-to-one fit using the empirical values from their Table II) and \cite{2018ApJS..237...21M} (a one-to-one fit using relations 2 \& 3 from their Table 10). For the planetary relation we show three published relations; \cite{2016ApJ...825...19W} (a probabilistic relation), Forecaster\footnote{\href{https://github.com/chenjj2/forecaster}{https://github.com/chenjj2/forecaster}} from \cite{2017ApJ...834...17C} (a probabilistic relation) and MRExo\footnote{\href{https://github.com/shbhuk/mrexo}{https://github.com/shbhuk/mrexo}} from \cite{2019arXiv190300042K} (a one-to-one fit).

For each tested method we computed the sum of the absolute values of the residuals between the method and the NASA data. The methods with the smallest total residuals were then chosen. Based on this test it was found that the stellar mass-radius relation was best approximated by the cubic fit whereas the planetary relation was best approximated using the MRExo relation. Therefore these are the mass-radius relations used in the simulation presented here.

\subsubsection{Instrumental noise}
\label{sec:Instrumental noise}

\begin{figure}
    {\includegraphics[width=\columnwidth]{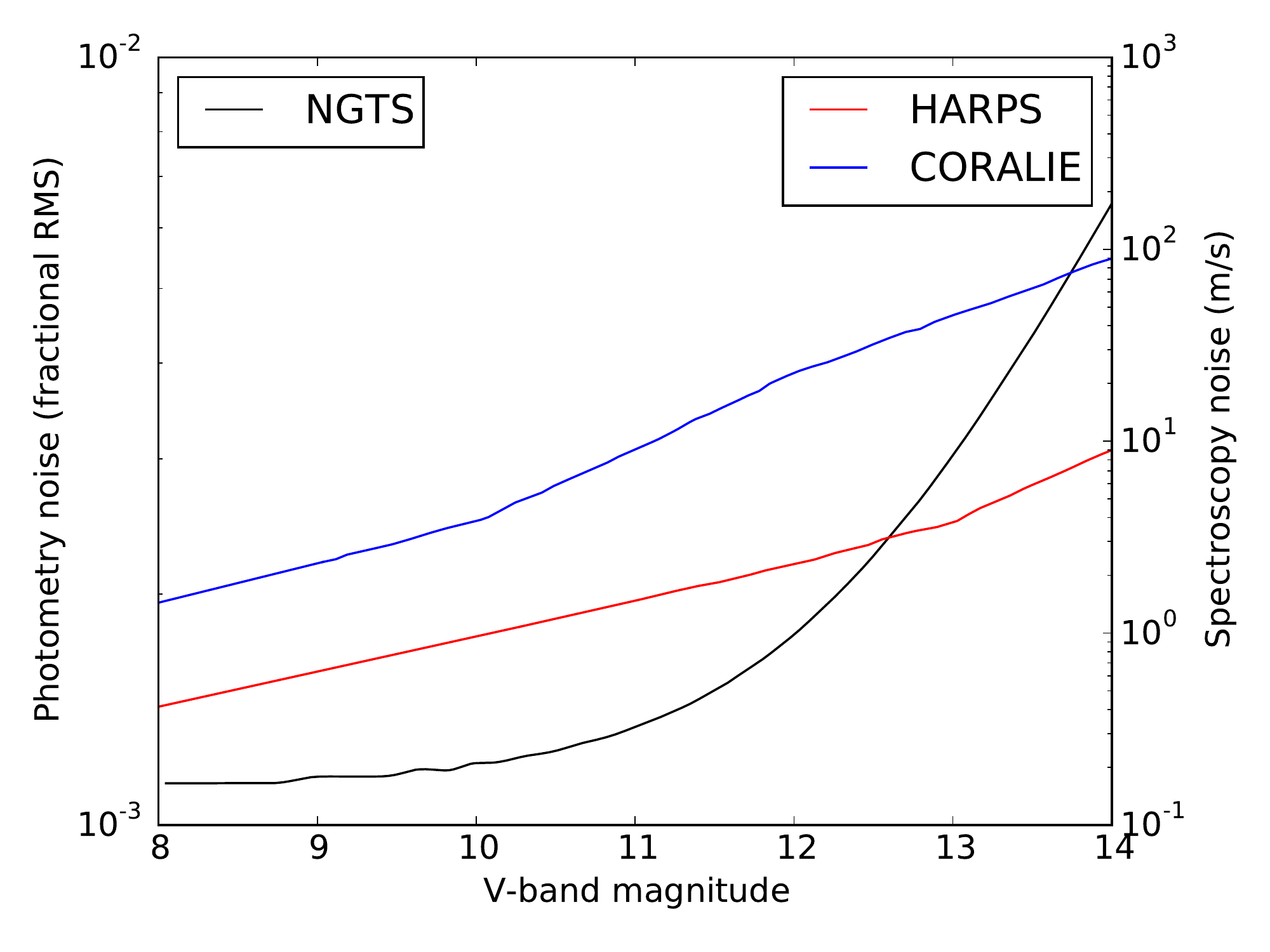}}
    \caption{Instrumental noise as a function of target magnitude for NGTS, HARPS and CORALIE (adapted from Figure 14 of \protect\cite{2018MNRAS.475.4476W} and Figure 10 of \protect\cite{2017MNRAS.465.3379G}).}
\label{fig:combined noise}
\end{figure}

Once the size of the photometric or spectroscopic signal has been calculated it is also necessary to determine the amount of noise that would be present on such a signal. Using this the SNR can then be calculated and it can be determined whether the signal would be observed. The levels of noise are instrument specific. The instruments employed in this simulation are NGTS 
for photometry and HARPS and CORALIE for spectroscopy. For each of these instruments the noise levels are functions of the magnitude of the host star. Figure \ref{fig:combined noise} shows the noise thresholds as a function of magnitude for these three instruments. The NGTS noise model (black) is adapted from Figure 14 of \cite{2018MNRAS.475.4476W} and the HARPS and CORALIE noise models (red and blue respectively) are adapted from Figure 10 of \cite{2017MNRAS.465.3379G}.

Using these data and the calculated signal sizes we can now determine the SNR for each instrument at each point in the parameter space to be explored.

\subsection{Follow-up times}
\label{sec:Follow-up time}

The simple calculation of SNR for each of the two methods is not sufficient to determine which method is the best follow-up technique. It is also pertinent to take into account the amount of time required for this follow-up. For example, a factor of 2 difference in SNR using each method is broadly irrelevant if the higher SNR method requires 100 times more observing hours. To account for this it we attempted to quantify the amount of time each method would require to accurately characterise each system.

\subsubsection{Photometry}
\label{sec:Photometry}

For photometry it was decided that a system was sufficiently followed-up when at least 2 additional transits (or partial transits) had been detected. Here we define this to mean at least two distinct nights in which at least an ingress or egress is seen during the hours of astronomical night. This means we require either the start or end of a transit ($T_c \pm T_{dur}/2$) to happen during the hours of darkness. The length of a night varies across the year but is, on average, $\sim$\,8 hrs \citep{2018A&A...619A.175C} so we use this value to avoid biasing our results by when the observations are begun. This calculation then requires an additional parameter, transit duration $T_{dur}$. Assuming an edge-on, circular orbit transit duration is given by

\begin{equation}
\label{eq:tdur}
    T_{dur} = \frac{P}{\pi}\arcsin\left({\frac{R_\star + R_p}{a}}\right),
\end{equation}

where symbols are as described above \citep{2010arXiv1001.2010W}. Using the calculated value for $T_{dur}$ together with the period it is straightforward to calculate how long until 2 transits are observed. However, this time will be affected by a number of random processes, specifically, what fraction of the period has elapsed when observations begin and what fraction of nights are adequate for photometric observations. Since there is some randomness involved in the calculation it makes sense to run the process multiple times, taking the average required observing time at the conclusion of the simulation. In this simulation we run the process 10 times (larger numbers of iterations were tested and found to have little effect on the results but a large increase in computing time) for each value of period, planetary radius and stellar radius (and therefore, transit duration). For each possible observed transit we apply a constraint assuming that 20\% of nights are inadequate for observations, thus transits which occur on these nights are missed. This fraction is based on a combination of bad weather effects, technical downtime and non-constant access to telescope time (see Figure 9 of \cite{2018MNRAS.475.4476W} for more details). As an approximation we assume this fraction is constant though it is most likely to vary throughout a year. Additionally, we assume that observations begin at a completely random point in the orbital phase.

Each simulation is run for a time of $10P$, where $P$ is the period of the system is question. 
If the simulation has not observed 2 transits within this time frame we set the follow-up time equal to $10P$ and move on. This has the effect of putting a lower bound on the follow-up time of some systems but reduces the computing time required sufficiently as to make the simulation tractable. As it turns out the vast majority of systems are followed-up before reaching this threshold. This fact, combined with the averaging of 10 runs means this limit has only a marginal impact on the results.

\subsubsection{Spectroscopy}
\label{sec:Spectroscopy}

For spectroscopy we define a system as sufficiently followed up when we can determine its period, RV amplitude and phase to within 5\% of true values from spectroscopy measurements alone. To model this we make the assumption that the orbit is circular and can therefore be fit with a sine curve of the form


\begin{equation}
\label{eq:sine}
    y = K\sin\left({\frac{2\pi}{P}x + \phi}\right) + \gamma.
\end{equation}


This assumes that the planetary signal we are searching for is the only cause of RV variations. In reality there may be other signals including additional planets or stellar variability. This assumption is a necessary simplification to achieve the general results of this simulation but it should be noted that spectroscopic follow-up times are generally a lower bound (especially for small amplitude signals) due to these additional effects.

To predict spectroscopic follow-up time $T_{spec}$ we simulate RV data points being taken and continue until the number of data points is sufficient to allow estimation of period and RV semi-amplitude to sufficient precision. We simulate RV observations in somewhat of a targeted strategy, assuming an observation every $n$ days where $n = \ceil[\Big]{\frac{P}{30}}$. As with photometry we assume that on 20\% of observing nights we cannot obtain data, accounting for weather, technical issues and higher-priority telescope targets. The effect of weather on spectroscopy is lessened compared to photometry but the instruments considered here are likely to have many programs running, reducing availability for specific follow-up. Thus the same 20\% was utilised. For each data point obtained we calculate its amplitude using equation \ref{eq:sine} with 
$\phi = \gamma = 0$. We then introduce a noise to the measurement drawn from a Gaussian distribution with $\mu$ equal to zero and $\sigma$ equal to the noise at the chosen magnitude from Figure \ref{fig:combined noise}.


Every time a new data point is added we create a Lomb-Scargle periodogram \citep{1976Ap&SS..39..447L,1982ApJ...263..835S} of the current data. If the highest power period is a match to the true period (to within 5\%) we then fold the data on this found period and attempt to fit it with a sine wave of the form shown in equation \ref{eq:sine} (we fit to the folded data as it was found that this method reduces the computational time required). Since the data is already folded on the correct period we can simplify equation \ref{eq:sine} by setting 
$P = 1$ and $\gamma = 0$. The fit utilises a least squares method to fit the equation and upholds the requirement for SNR $\geq 3.0$. The fit then returns parameter $K$ which is the RV semi-amplitude of the signal, and $\phi$ which is the phase of the orbit. If $K$ and $\phi$ are within 5\% of the true values we consider the signal recovered and record the time taken. Otherwise we simulate a new data point and repeat the process.

As for photometry, each simulation is run for a time of $10P$ where $P$ is the period of the system is question. 
If the simulation has not solved the orbit to 5\% within this time frame we set the follow-up time equal to $10P$ and move on. Once again we find that the vast majority of systems are followed-up before reaching this threshold resulting in only a marginal impact on the results. We make two additional requirements for spectroscopic follow-up. We require at least 5 valid data points and assume the systems cannot be characterised before it has been observed for at least half an orbital period.

\subsubsection{Instrument availability}
\label{sec:Instrument availability}

A factor that requires more discussion is that of telescope availability for the type of follow-up envisioned by this simulation. A photometric instrument like NGTS can be operated in a dedicated fashion. Because of this the main impacts that would reduce the instrument time that can be spent on this follow-up are external factors such as weather and instrument failure. These factors have been combined into an average of 20\% of nights being removed as photometrically feasible, as mentioned in Section \ref{sec:Photometry}. For the spectroscopic instruments discussed here (and particularly for HARPS) this is less feasible. Observers wishing to use these instruments must contend with potentially disrupted schedules due to the popularity of the instrument and the process of user queues. This is a hard effect to model, especially into the future when schedules are unknown. Our solution to this problem is to remove a random fraction of nights to simulate these falling into the gaps of the schedules. The benefit of a spectrograph is that they are less affected by bad weather, that is, conditions can be sufficient for spectroscopic observations but not for photometric ones. Because of these two factors combined, and informed by past user queues for these instruments, we choose to use the same fraction of 20\% of nights as being unsuitable for spectroscopy.

\section{Results}
\label{sec:Results}
 
The results presented below all show the case for an $V = 11$ magnitude host star.

\subsection{SNR results}
\label{sec:SNR results}

\begin{figure}

    \begin{subfigure}[HARPS/NGTS SNR
    \label{fig:NGTS/HARPS SNR}]
    {\includegraphics[width=\columnwidth]{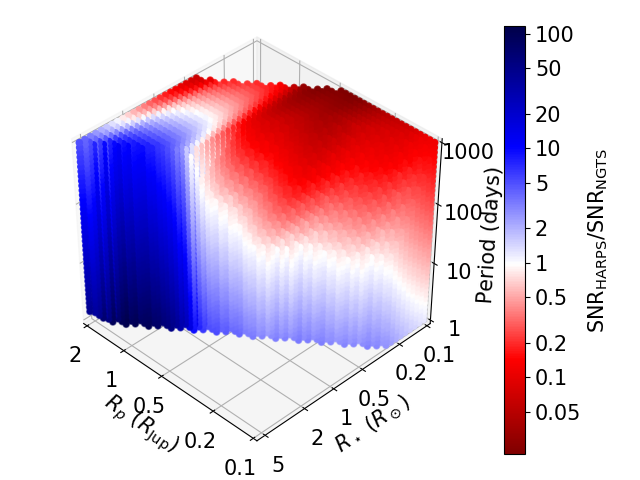}}
    \end{subfigure}
    \begin{subfigure}[HARPS/NGTS SNR plus
    \label{fig:NGTS/HARPS SNR plus}]
    {\includegraphics[width=\columnwidth]{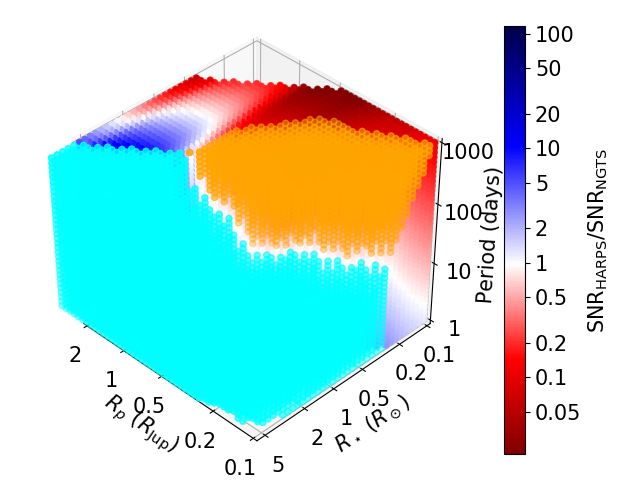}}
    \end{subfigure}
    \caption{SNR ratio in NGTS/HARPS parameter space. The gradated colour area indicates the region accessible to NGTS and HARPS with the colour indicating the ratio of their respective SNR values (bluer where HARPS is better, redder where NGTS is better). The solid orange region in Figure (b) indicates the region accessible to NGTS only and the solid cyan region in Figure (b) indicates the region accessible to HARPS only. Empty regions are unphysical or inaccessible systems.}

\label{fig:NGTS/HARPS SNR ratios}
\end{figure}

Figure \ref{fig:NGTS/HARPS SNR} shows a plot of the area of $R_\star$, $R_p$, $P$ parameter space that can be observed by NGTS and HARPS with an SNR $\geq3.0$. The colour bar on this plot corresponds to the ratio of the predicted SNR that could be reached with NGTS and HARPS with values $>1$ being points for which HARPS would reach a better SNR and values $<1$ indicating points for which NGTS would reach a better SNR. From this plot and the values in Table \ref{tab:different mags} it can be seen that approximately 48\% of the physically realistic phase space can be accessed by both NGTS and HARPS. This 48\% is then split approximately evenly (42:58) between points for which HARPS would achieve a higher SNR and points for which NGTS would be better. Figure \ref{fig:NGTS/HARPS SNR plus} shows the same range of points as Figure \ref{fig:NGTS/HARPS SNR} but this time includes the additional two areas that correspond to the regions of parameter space that only 1 instrument can reach with the required SNR $\geq3.0$ (NGTS in orange and HARPS in cyan). The plot shows that there is a larger fraction of parameter space accessible to only HARPS (22\%) than there is to only NGTS (5\%). Key from these plots is that it is the longer period, smaller planet systems which are best followed by NGTS and the shorter period, larger planet systems that are best followed by HARPS. The empty regions of this plot correspond to the parameter space that is associated with unphysical systems (for example, planet larger than the star, or orbital separation less than stellar radius) as well as those regions for which neither NGTS nor HARPS can reach at sufficient SNR.

\begin{figure}

    \begin{subfigure}[CORALIE/NGTS SNR
    \label{fig:NGTS/CORALIE SNR}]
    {\includegraphics[width=\columnwidth]{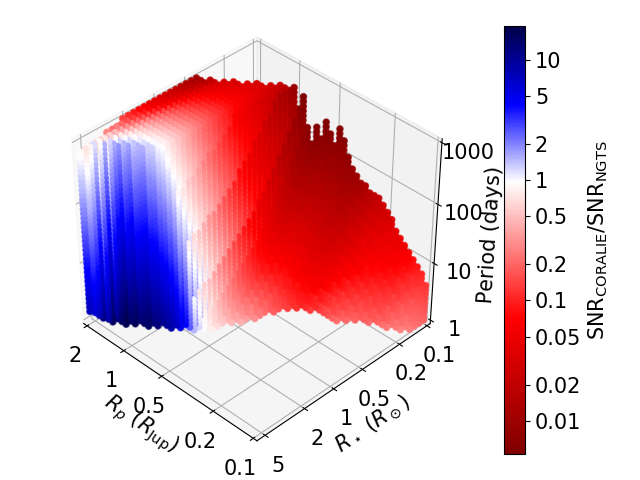}}
    \end{subfigure}
    \begin{subfigure}[CORALIE/NGTS SNR plus
    \label{fig:NGTS/CORALIE SNR plus}]
    {\includegraphics[width=\columnwidth]{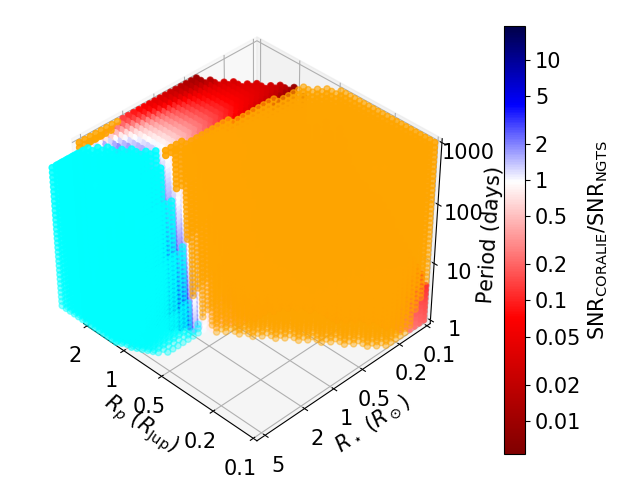}}
    \end{subfigure}
    \caption{SNR ratio in NGTS/CORALIE parameter space. The gradated colour area indicates the region accessible to NGTS and CORALIE with the colour indicating the ratio of their respective SNR values (bluer where CORALIE is better, redder where NGTS is better). The solid orange region in Figure (b) indicates the region accessible to NGTS only and the solid cyan region in Figure (b) indicates the region accessible to CORALIE only.}

\label{fig:NGTS/CORALIE SNR ratios}
\end{figure}

Figures \ref{fig:NGTS/CORALIE SNR} and \ref{fig:NGTS/CORALIE SNR plus} show the results of the SNR distribution again, this time comparing NGTS to CORALIE. As expected, there is a smaller region of space accessible to both instruments (27\%) due to CORALIE's higher noise values, however the overall trend is the same with the same regions of parameter space lending themselves to photometry and spectroscopy. In Figure \ref{fig:NGTS/CORALIE SNR plus} it is clear that there is now significantly more parameter space that can be accessed by NGTS only (25\%) than by CORALIE only (8\%), a simple result of CORALIE being unable to reach as many low radial velocity signal systems as HARPS can.

\subsection{Follow-up time results}
\label{sec:Follow-up time results}

\begin{figure}

    \begin{subfigure}[NGTS follow-up time
    \label{fig:NGTS wait time}]
    {\includegraphics[width=\columnwidth]{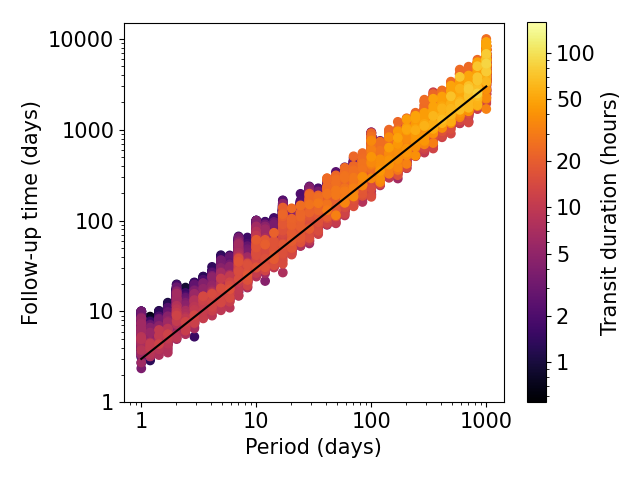}}
    \end{subfigure}
    \begin{subfigure}[HARPS follow-up time
    \label{fig:HARPS wait time}]
    {\includegraphics[width=\columnwidth]{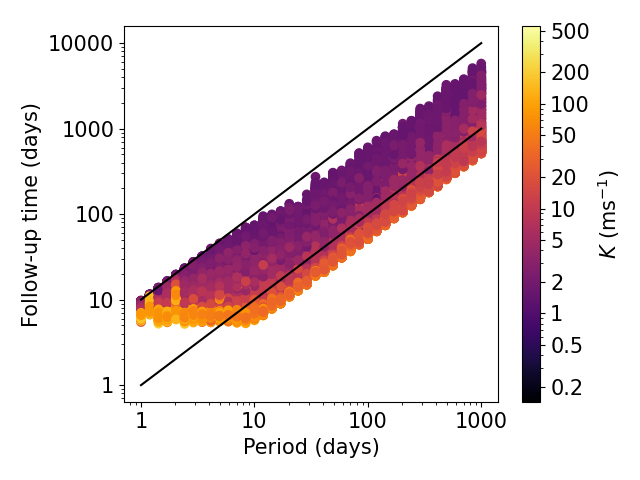}}
    \end{subfigure}
    \begin{subfigure}[CORALIE follow-up time
    \label{fig:CORALIE wait time}]
    {\includegraphics[width=\columnwidth]{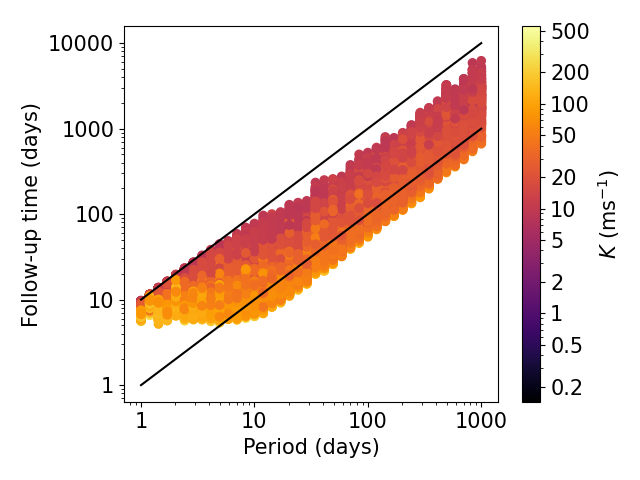}}
    \end{subfigure}
    \caption{Predicted follow-up time as a function of period for NGTS, HARPS and CORALIE. For NGTS a colour bar denoting transit duration is shown and for HARPS and CORALIE a colour bar denoting RV amplitude is used. Additionally, Figure (a) includes a linear $y = 3x$ trend and Figures (b) and (c) include linear $y = x$ and $y = 10x$ trends to aid the eye. Each plot shows all points accessible to that instrument.
    }

\label{fig:wait times}
\end{figure}

Figures \ref{fig:NGTS wait time}, \ref{fig:HARPS wait time} and \ref{fig:CORALIE wait time} show plots of period against predicted follow-up time for NGTS, HARPS and CORALIE respectively. These follow-up times indicate the time until either two transits are observed (photometry) or period, amplitude and phase of RV signal are constrained to 5\% or better (spectroscopy). These follow-up times are predicted using the methods set out in sections \ref{sec:Photometry} and \ref{sec:Spectroscopy}. The plots additionally show some linear trends to aid the eye. Figure \ref{fig:NGTS wait time} includes a trend at $y = 3x$ and Figures \ref{fig:HARPS wait time} and \ref{fig:CORALIE wait time} include trends at $y = x$.

For photometry it can be seen that the relation between period and follow-up time is approximately linear with the time required for two transits being approximately equal to 3 times the period. It is also seen that, for periods $\gtrsim20$\,days, this is relatively independent of the transit duration, shown in the colour bar. For shorter periods however, we do see a dependence, with shorter transit durations requiring more follow-up time.

For spectroscopy the trend is linear at larger periods ($P\gtrsim10$ day) for all $K$ but flattens off at shorter periods for large $K$ values. This however is a direct result of our requirement for at least 5 valid spectroscopic points. For spectroscopy there is a dependence of RV amplitude with lower amplitude signals requiring more time to properly constrain as seen in the colour bar. Additionally, by comparing Figures \ref{fig:HARPS wait time} and \ref{fig:CORALIE wait time} it can be seen that, when comparing a system with the same period and RV amplitude, it takes CORALIE longer to constrain a system than HARPS, as is expected from the worse noise performance of CORALIE compared to HARPS. Each plot shows all points accessible to that instrument.

\begin{figure}

    \begin{subfigure}[NGTS/HARPS follow-up times
    \label{fig:NGTS/HARPS wait times}]
    {\includegraphics[width=\columnwidth]{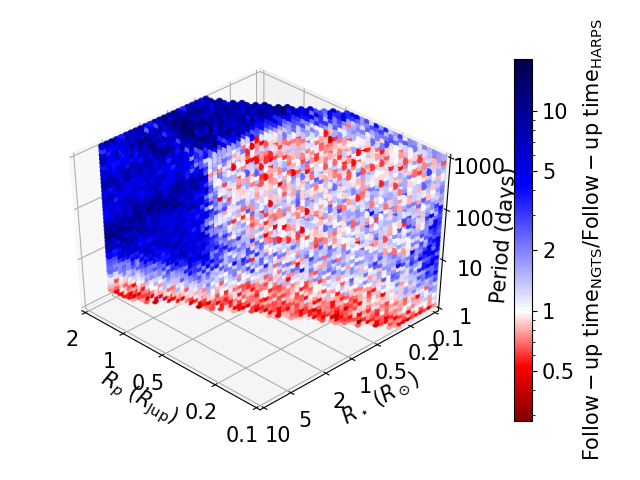}}
    \end{subfigure}
    \begin{subfigure}[NGTS/CORALIE follow-up times
    \label{fig:NGTS/CORALIE wait times}]
    {\includegraphics[width=\columnwidth]{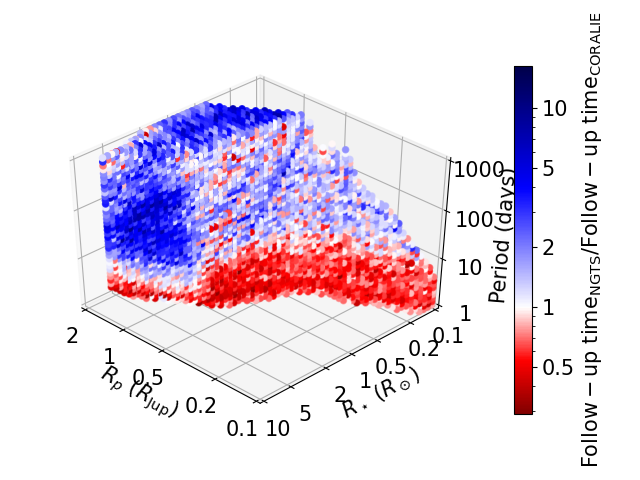}}
    \end{subfigure}
    \caption{Follow-up time ratios in $P$, $R_p$, $R_\star$ parameter space. Plots (a) and (b) show comparisons of NGTS/HARPS and NGTS/CORALIE respectively. For each plot only the region for which both instruments achieve SNR $\geq3.0$ is shown. The colour bars denote the follow-up time ratios with values $<1$ indicating a shorter follow-up time for photometry and values $>1$ indicating shorter follow-up times for spectroscopy.}

\label{fig:wait time ratios}
\end{figure}

Figures \ref{fig:NGTS/HARPS wait times} and \ref{fig:NGTS/CORALIE wait times} show the same regions of parameter space as shown in Figures \ref{fig:NGTS/HARPS SNR} and \ref{fig:NGTS/CORALIE SNR}. This time however, the plots are coloured by a ratio of the predicted follow-up times for NGTS \& HARPS and NGTS \& CORALIE respectively. 

The main point of interest from these plots is that there are regions of parameters space more amenable to follow-up by both methods. For the very shortest period planets ($\lesssim$\,10\,d) photometry is quicker at follow-up, most likely due to the additional requirements for at least 5 spectroscopic points. For longer periods spectroscopy is quicker for follow-up of those systems with larger planets whereas photometry can be faster for smaller planet systems. The ratio of follow-up times is broadly independent of stellar radius. The same general pattern is seen in both figures with the main difference being a slight preference for photometry in Figure \ref{fig:NGTS/CORALIE wait times} compared to Figure \ref{fig:NGTS/HARPS wait times} as a result of HARPS being quicker to constrain the same system than CORALIE (see Figures \ref{fig:HARPS wait time} and \ref{fig:CORALIE wait time}). Not shown in these plots are the regions of parameter space that only one instrument can access at the required level of SNR. The instrument that can quicker constrain a given system in these regions is broadly irrelevant as only one instrument would actually be able to confidently measure the required signal.


An additional point that must be mentioned in regards to follow-up time is the fact that stars are only seasonally observable from the ground. To this end, the total follow-up times may be longer than those shown here, particularly for systems with periods above $\sim$\,180\,days. Since the three instruments considered here are all withing geographic proximity to one another (all being located in northern Chile) this affect should impact all observations equally. Because the point of interest in this simulation is the relative follow-up time between the different methods the follow-up times are not extended since there would be no change to the relative times. However this is an effect that should be considered when planning real observations.

\subsection{Other magnitudes}
\label{sec:Other magnitudes}

The results and plots presented thus far in this paper have looked at a host star of $V=11$ magnitude. These results however, will not be identical for different magnitudes as the noise profiles of the three instruments varies (see Figure \ref{fig:combined noise}). For this paper we reran our analysis at $V=8$ and $V=14$. The broad patterns in the results are consistent with similar regions of parameter space preferring photometry and spectroscopy as for the $V=11$ case. The main difference between the three cases is the amount of parameter space which is or isn't available to the three instruments. Where  multiple instruments are viable the outcomes are broadly unchanged. Table \ref{tab:different mags} shows how the accessible regions of parameter space vary with magnitude.

\begin{table}
 \caption{Accessible parameter space as a function of host magnitude. Given as a percentage of the full physically realistic range of points considered (57847 points).}
 \label{tab:different mags}
 \begin{tabular}{lccc}
  \hline
  Magnitude ($V$-band) & 8 & 11 & 14\\
  \hline
  NGTS (total) & 54 & 53 & 37\\
  NGTS (only) & 0 & 5 & 15\\
  HARPS (total) & 94 & 70 & 35\\
  HARPS (only) & 19 & 14 & 11\\
  CORALIE (total) & 71 & 35 & 7\\
  CORALIE (only) & 0 & 0 & 0\\
  NGTS and HARPS & 54 & 48 & 22\\
  NGTS and CORALIE & 49 & 27 & 6\\
  NGTS and HARPS and CORALIE & 49 & 27 & 6\\
  \hline
 \end{tabular}
\end{table}

As this table shows, and as expected, increasing host magnitude leads to a reduction in the volume of parameter space available to each instrument. This is a natural consequence of the increased noise that accompanies dimmer targets. The table also confirms that there is no area of parameter space accessible to CORALIE and not HARPS. Once again this is as expected since HARPS has a lower noise floor than CORALIE at all magnitudes.

There is an additional preference towards spectroscopy for brighter targets. Brighter stars lead to lower noise floors an higher SNR value for all instruments but spectroscopic follow-up time is also affected. The search for additional transits is not affected by host magnitude since only transits with $SNR\geq3.0$ are considered. For spectroscopy however the noise value is built into the simulation of data and fitting routines. These should then be more efficient for brighter hosts leading to a reduced follow-up time for spectroscopy. For a system that can be observed at $SNR\geq3.0$ at both $V=8$ and $V=11$ the follow-up time for photometry is unaffected whereas the follow-up time for spectroscopy is reduce for the brighter target.



\subsection{Online repository}
\label{sec:Online repository}

The results presented here in Figures \ref{fig:NGTS/HARPS SNR ratios}, \ref{fig:NGTS/CORALIE SNR ratios}, \ref{fig:wait time ratios} and \ref{fig:weighted} are attempts to display 3D information in a 2D medium. This is obviously non-ideal so we provide a link to an online repository containing the 3D data and plotting script required to create these figures. The details are included to recreate and manipulate all plots for $8^{\rm th}$, $11^{\rm th}$ \& $14^{\rm th}$ $V$-band magnitude hosts. The repository is to be found at \href{https://github.com/BenCooke95/SpecPhot}{https://github.com/BenCooke95/SpecPhot}. This repository also contains the codes necessary to reproduce the simulation discussed is this paper.

\section{Discussion}
\label{sec:Discussion}

\subsection{SNR}
\label{sec:SNR discussion}

The methods used to calculate SNR presented in this paper are fairly simplistic but are a realistic approximation. In reality, of course, SNR can be increased by taking more measurements and binning data points. This can then beat down the noise and allow smaller signals to show though. In this paper however we are interested in the broader picture for which considering SNR as the simple ratio of signal size to noise size is sufficient.

To explain the behaviour of our SNR ratio with $P$, $R_\star$ and $R_p$ we can look at equations \ref{eq:phot signal}, \ref{eq:spec signal} and \ref{eq:a}. Combining these equations to find the ratio of the signals gives us the relation

\begin{equation}
\label{eq:specphot ratio}
    \frac{K}{\delta} \propto \frac{M_pR_\star^2}{R_p^2(M_\star+M_p)^\frac{2}{3}P^{\frac{1}{3}}}.
\end{equation}

Taking $M_\star \propto R_\star^3$ (see Figure \ref{fig:mass-radius stars}, the relation used in this paper) and $M_p \propto R_p^3$ (see Figure \ref{fig:mass-radius planets}, an approximation to the relation used in this paper) and assuming $R_\star^3 \gg R_p^3$ we find that

\begin{equation}
\label{eq:specphot ratio simple}
    \frac{K}{\delta} \propto \frac{R_p}{P^{\frac{1}{3}}}.
\end{equation}

This relation now explains our results well. We see clearly that longer period systems favour better photometry SNR than spectroscopy in accordance to the $P^{-\frac{1}{3}}$ term. Additionally, our SNR ratio is broadly independent of stellar radius but generally favours spectroscopy for larger radius planets. Relaxing the approximations reasoned here and allowing for the more realistic planet mass-radius relation used explains the distributions seen in Figures \ref{fig:NGTS/HARPS SNR ratios} and \ref{fig:NGTS/CORALIE SNR ratios}.

\subsection{Follow-up time}
\label{sec:Follow-up time discussion}

There are many possible definitions of follow-up time, or how long before a system is sufficiently characterised. In this paper we have defined a successful follow-up campaign as one which either observes two additional transits (photometry) or measures period, RV amplitude and phase to 5\% or better (spectroscopy). In reality, the exact conditions required for a successful follow-up may change in a system dependent way but we consider the criteria stated here as reasonable baselines.

For photometry we see that the follow-up time is approximately equal to $3P$ and largely independent of other parameters. This makes sense since the frequency with which transits occur is $1/P$. A longer transit duration will help to increase the chance that at least part of a transit occurs during a good observing night but this is a small effect compared to the decreasing frequency of events towards longer periods. The fact that the follow-up time is not always $\leq 2P$ is due to the fact that we require part of a transit during a night of good observing quality.

The follow-up time to period relation for spectroscopy is more detailed. For longer period systems the follow-up time is usually $\gtrsim P$ with smaller amplitude signals requiring longer to follow up. This implies that we generally require close to or above 100\% phase coverage to accurately measure period and RV amplitude for long period systems. Within this regime higher amplitude signals are characterised with fewer data. The relation is approximately 1:1 for $K$ values of $\sim\,20\rm\,ms^{-1}$ and $\sim\,50\rm\,ms^{-1}$ for HARPS and CORALIE respectively but is closer to 5:1 for $K$ values of $\sim\,4\rm\,ms^{-1}$ and $\sim\,10\rm\,ms^{-1}$ for HARPS and CORALIE respectively. For shorter period systems, less than around 10\,days, the linear relation begins to break down (at least for larger $K$). This is due to our requirement for $\geq$5 data points before attempting characterisation. 
Even in this regime the effect of larger amplitude signals is still seen, though to a lesser degree due to the clumping of points between 5 and 10 days.

As an additional factor to consider we see that when the period is an integer day the follow-up time can spike to infinity due to every transit occurring during the day (for photometry) or only a limited amount of phase coverage being possible (for spectroscopy). This is a known issue for ground-based observations and the difficulty of characterisation leads to a dearth of integer day periods, not representative of the true distributions.


\subsection{Weighting}
\label{sec:Weighting}

\begin{figure}

    \begin{subfigure}[HARPS/NGTS weighted values
    \label{fig:NGTS/HARPS weighted}]
    {\includegraphics[width=\columnwidth]{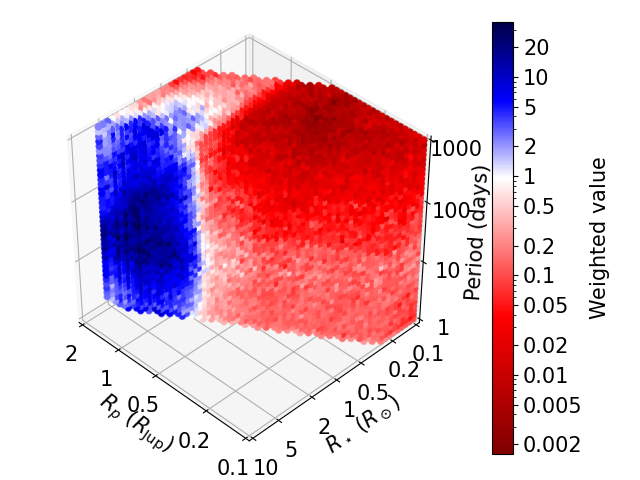}}
    \end{subfigure}
    \begin{subfigure}[CORALIE/NGTS weighted values
    \label{fig:NGTS/CORALIE weighted}]
    {\includegraphics[width=\columnwidth]{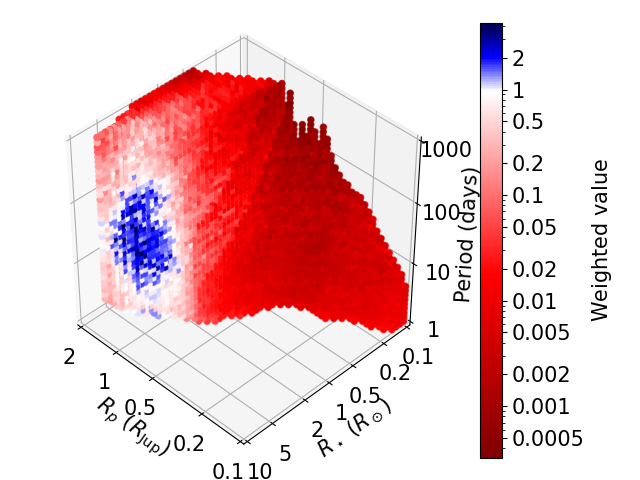}}
    \end{subfigure}
    \caption{Weighted combinations of SNR and follow-up time. These plots are a combination of the distributions in Figures \ref{fig:NGTS/HARPS SNR ratios}, \ref{fig:NGTS/CORALIE SNR ratios} and \ref{fig:wait time ratios} weighted by a factor of $\frac{1}{12}$. The colour bar shows the weighted value with values $<1$ indicating a preference for photometry (NGTS) and values $>1$ indicating a preference for spectroscopy (HARPS or CORALIE).}

\label{fig:weighted}
\end{figure}

To be able to rank individual parameter combinations in terms of SNR and follow-up time requires combining both terms into a single value. Additionally, for a more realistic comparison we must also take into account the availability of the two methods being compared in terms of telescope time. Finding the appropriate weighting that takes these factors into account is non-obvious and, ideally, bespoke to the needs and resources of a specific follow-up effort or research group. For the purposes of this paper we combine the factors in a simplistic way. We simply multiply the ratio of SNR values and the ratio of follow-up times. This leads to a single number for each set of $R_\star$, $R_p$ and $P$ values with values $>1$ meaning spectroscopy is preferred and values $<1$ meaning photometry is preferred. Additionally, we multiply by a  factor of $\frac{1}{12}$ to reflect that each of HARPS and CORALIE comprise of a single instrument whereas NGTS contains 12 individually operable telescopes \citep{2018MNRAS.475.4476W}. Figure \ref{fig:weighted} shows the weighted distributions for NGTS \& HARPS and NGTS \& CORALIE.

Comparing Figure \ref{fig:weighted} to Figures \ref{fig:NGTS/HARPS SNR ratios} and \ref{fig:NGTS/CORALIE SNR ratios} we see that the weighting has caused noticeable changes, most significantly for the NGTS/HARPS comparison. For NGTS \& HARPS we see that the area of the distribution more favourable to photometry has increased compared to Figure \ref{fig:NGTS/HARPS SNR ratios}, especially towards low periods. The factor of $\frac{1}{12}$ would lead to an increased preference for photometry across the distribution but small period range is more affected due to the increased follow-up time spectroscopy has in this parameter range (seen in Figure \ref{fig:wait times}). The NGTS/CORALIE distribution is similarly affected with an enhancement of photometry towards low periods. The distribution is less obviously difference due to the smaller region of parameter space accessible to CORALIE and the already weaker SNR and follow-up times when compered to HARPS.

\subsection{Importance of results}
\label{sec:Importance of results}

Follow-up of exoplanetary systems, especially poorly characterised ones, is a detailed undertaking that can involve different criteria based on required outcome and potential resources. The myriad specifics that could potentially be involved are too numerous to explore in a paper such as this so we make some simplifying assumptions and choose to focus on the broadly applicable results. Even at this level of detail, however not all of the results presented here are obvious. For example, the overall relation describing the ratio of spectroscopic to photometric SNR (equation \ref{eq:specphot ratio simple}) can be simply found by making some assumptions to the common transit and radial velocity equations but the more detailed structure present in Figures \ref{fig:NGTS/HARPS SNR ratios} and \ref{fig:NGTS/CORALIE SNR ratios} is not straight forward to predict. The precise dividing line between the two regimes cannot be intuited but may be vital when arguing for one method over another. The follow-up time data is not necessarily obvious either. The fact that short period planets can be quickly confirmed using photometry is easy to explain (multiple transit opportunities within only a few days) but the fact that there are systems up to even the longest periods for which photometry is the quicker follow-up method is surprising.

A simulation like this is a good starting point when attempting to garner time on a particular instrument for follow-up by showing that other methods have been considered and that the method you are targeting is appropriate for your specific case. Securing time on spectroscopy or photometry instruments can be difficult and results like these, showing that one method is preferred or more efficient, are useful evidence.

\section{Conclusions}
\label{sec:Conclusions}

The results presented in this paper show how the effectiveness of using photometry or spectroscopy for exoplanet follow-up changes as a function of $R_\star$, $R_p$ and $P$. In terms of purely measuring the SNR value of a detection we see that photometry is favoured for longer period systems and for smaller radius planets whereas spectroscopy is favoured for shorter period systems and larger radius planets. Generally both methods perform comparably regardless of stellar radius. In terms of follow-up time, we find that photometry usually follows-up a system faster for small planet systems whereas spectroscopy is faster for larger planets. The region where this becomes less of a factor is towards shorter period systems where follow-up times generally prefer photometry due to the spectroscopic methods departure from a linear relation.

Because of this dependence on period it can be difficult to choose the best method when the period of the system is uncertain (the case for monotransits). As mentioned before, the shape of the discovery transit can inform the period with an accuracy up to 10\%, generally this is accurate enough to help determine the chosen follow-up method (especially if relying on the weighted ratios as in Figure \ref{fig:weighted} which are less dependent on period). However we mention here that extra care must be taken for systems with poorly constrained periods.

To choose the best method with which to follow-up these systems requires taking into account both criteria, SNR and follow-up time, with an appropriate weighting based on the specific goals of the follow-up efforts. For borderline signal cases one should most likely weight the SNR ratio higher, whereas for obvious but scientifically important systems follow-up time is likely to be more important. Additionally, both methods must be considered accounting for availability of telescope time, a shorter amount of required spectroscopic time may be a moot point if photometric time is more readily available. We make an estimation of a typical weighting based on the larger number of NGTS telescopes but bespoke weighting based on specific research goals and resources would doubtless be more realistic.

\section*{Acknowledgements}
\label{sec:Acknowledgements}

We thank the anonymous referee for their comments which have helped to improve this paper. BFC acknowledges a departmental scholarship from the University of Warwick. DP acknowledges support through a Merit Award from The Royal Society and from the Science and Technology Facilities Council (STFC) ST/P000495/1.




\bibliographystyle{mnras}
\bibliography{specphot} 





\bsp	
\label{lastpage}
\end{document}